\documentclass{article}
\usepackage{amsmath,amssymb}
\title{The Preferred Frame and Poincar\'e Symmetry\thanks{Submitted
to the Proceedings of the XXI International Colloquium on
Group Theoretical Methods in Physics, 15--20 July,
Goslar, Germany}}
\author{
Jakub Rembieli\'nski
\thanks{{\it E-mail address\/}: jaremb@mvii.uni.lodz.pl,
jaremb@krysia.uni.lodz.pl}\ ,
Pawe{\l}
Caban\thanks{{\it E-mail address\/}: caban@mvii.uni.lodz.pl}
\vspace{1cm}\\
Department of Theoretical Physics, University of {\L}\'od\'z\\
Pomorska 149/153, 90--236 {\L}\'od\'z, Poland}
\date{December 1996}            

\newcommand{\nad}[1]{\mbox{\smash{\oalign{$#1$ \crcr \hidewidth 
$\mathchar"017E$ \hidewidth}}}}

\begin{document}
\maketitle
\begin{abstract}
In this paper we describe a covariant canonical formalism
for a free time-like (massive) as well as space-like
(tachyonic) particle in the framework of nonstandard
synchronization scheme. In this scheme one is able to
introduce absolute causality {\em without} breaking the
Poincar\'e invariance.
\end{abstract}

\section{Introduction}
As is well known, special relativity, irrespectively of its great
success in local description of the reality, has a number of
disadvantages. They are related to the Minkowski space-time
notion rather than to the  very
well experimentaly supported Poincar\'e symmetry.
The main difficulties are
connected with the absence of covariant canonical formalism
and the lack of dynamics for systems of relativistic particles
\cite{Sund}. Even more serious problems arise on the quantum
ground where in fact no fully consistent relativistic
quantum mechanics for systems with finite degrees of freedom exists.
In particularthere is  no covariant notion of localizability and
relativistic position operator \cite{Sund,Bacry}.
In this paper we briefly describe another point of view,
preserving Poincar\'e symmetry but changing the space-time notion.
We apply this formalism to the simplest physical
system---free particle and show that it is
possible to introduce covariant
canonical formalism (Poisson structure) for both time-like
and space-like particles. In the papers \cite{Rem1,CR1,CR2,CabRem}
these ideas are applied
to elaborate the
hypothesis of the tachyonic neutrino as well as to introduce
a covariant notion of localizability on the quantum level.

\section{Preliminaries}
The main idea is based on two well known facts: $(i)$  definition of
a coordinate time depends on the synchronisation scheme; $(ii)$ 
synchronisation scheme is a convention, because no experimental procedure
exists which makes it possible to determine the one-way velocity
of light without use of superluminal signals
\cite{Jammer,Will,MansSexell}.
Therefore there is a freedom
in the definition of the coordinate time. The standard choice is the
Einstein--Poincar\'e (EP) synchronisation with the one-way light velocity
isotropic and constant. This choice leads to the extremely simple form of
the Lorentz group transformations but the EP coordinate time allows
a covariant causality for time-like and light-like trajectories only.
We choose a  different synchronisation, namely that  of
Chang--Tangherlini (CT), preserving invariance of the notion of the
instant-time hyperplane \cite{Rem1,Rem2}.
In this synchronisation scheme the
notion of causality is universal and space-like trajectories are
physically admissible too. The price is the more complicated form
of the Lorentz transformations incorporating transformation rules
for velocity of distinguished reference frame (preferred frame).
{\bf
The EP and CT descriptions are entirely equivalent if we restrict
ourselves to time-like and light-like trajectories;
however a consistent description of tachyons is possible only
in the CT scheme.}
A very important consequence is that if tachyons exist then the relativity
principle is broken,  i.e. there exists a preferred frame of reference,
however the Lorentz symmetry is   preserved.

The proper framework to this construction is the bundle of
Lorentzian frames; the base space is simply the space of velocities
of these frames with respect to the preferred frame.
For this reason the transformation law for coordinates incorporates
the velocity of distinguished frame. The preferred frame can be locally
identified with the comoving frame in the expanding universe
(cosmic background radiation frame) i.e. the reference frame of
the privileged observers to whom the universe appears isotropic
\cite{Weinberg}.

To be concrete the Lorentz group transformations in the mentioned
bundle of frames have the following form \cite{Rem1}:
 \begin{equation}
 \left\{\begin{array}{l}
 x^{\prime}=D(\Lambda,u)x \\
 u^{\prime}=D(\Lambda,u)u 
 \end{array}\right. \label{p1}
 \end{equation}
where for rotations $D(R,u)$ has the standard form while for boosts
it reads
 \begin{equation}\label{D:Wu}
 D(W, u) = \left(\begin{array}{c|c} \frac{1}{W^0} & 0 \\ \hline
 -\vec{W} & I +\frac{\vec{W} \otimes \vec{W}^{\mathrm{T}}}%
 {\left(1 + \sqrt{1 + (\vec{W})^2}\right)}
 -\vec{W} \otimes \vec{u}^{\mathrm{T}} u^0 \end{array}\right).
 \end{equation}
Here $W^{\mu}$ is the four-velocity of $(x^{\prime})$ frame
as seen by an observer in the frame $(x)$ while $u^{\mu}$ is the
four-velocity of the privileged frame as seen from the frame $(x)$.
Notice that the time coordinate is rescalled by a positive
factor only. The transformations (\ref{p1}) leaves
invariant the metric form
 \begin{equation}
 ds^2=g_{\mu\nu}(u)\,dx^{\mu}\,dx^{\nu}  \label{p3}
 \end{equation}
with
 \begin{equation}\label{g:u}
 \left[g_{\mu\nu}(u)\right] = \left(\begin{array}{c|c} 1 & u^{0}
 \vec{u}^{\mathrm{T}} \\[1ex] \hline u^{0} \vec{u} & 
 -I + \vec{u} \otimes \vec{u}^{\mathrm{T}} (u^{0})^{2}
 \end{array}\right).
 \end{equation}
Interrelation with coordinates in the EP
synchronization $(x_{E}^{\mu})$ is given by
 \begin{equation}
 x_{E}^{0}=x^0+u^0\vec{u}\vec{x},\qquad \vec{x}_E=\vec{x}.
 \label{p4}
 \end{equation}
However the corresponding interrelations
between velocities $\vec{v}_E$ and $\vec{v}$ obtained from (\ref{p4})
are singular for superluminal velocities.

\section{Covariant canonical formalism for the free time-like
(massive) particle}

Let us consider in detail the case of a free particle
associated with a time-like geodesics.
The corresponding action $S$ is of the form
 \begin{equation}\label{Sb}
 S_{12} = -{m} \int^{\lambda_2}_{\lambda_1} \sqrt{ds^2}
 \end{equation}
where the square of the time-like line element
 \begin{equation}\label{ds2<0b}
 ds^2 = g_{\mu\nu}(u) \frac{dx^\mu}{d\lambda}
 \frac{dx^\nu}{d\lambda} d\lambda^2 > 0
 \end{equation}
and the continuous affine parameter $\lambda$ is defined along
the trajectory as monotonically increasing as one proceeds
along the curve in a fixed direction.

The equations of motion are obtained by means of the
variational principle and reads
 \begin{equation}\label{d/dlb}
 \frac{d}{d\lambda} \left(
 \frac{\dot{x}^\mu}{\sqrt{g_{\mu\nu}(u) \dot{x}^\mu \dot{x}^\nu}}
 \right) = 0 
 \end{equation}
with $\dot{x}^\mu = \frac{dx^\mu}{d\lambda} \equiv w^\mu$.
Now, we are free to take the path parameter as
$d\lambda = \sqrt{ds^2}$, so the four-velocity $w^\mu$ satisfies
 \begin{equation}\label{w2b}
 w^2 = g_{\mu\nu}(u) w^\mu(u) w^\nu(u) = 1
 \end{equation}
and consequently
 \begin{equation}\label{w.b}
 \dot{w}^\mu = \ddot{x}^\mu = 0.
 \end{equation}
Defining velocity in a standard way as 
$\vec{v} = \frac{d\vec{x}}{dx^0} = \frac{\vec{w}}{w^0}$
we can identify the Lagrangian of a free particle related to
the action (\ref{Sb}); by means of the formulas
(\ref{p3}), (\ref{g:u}) we have
 \begin{equation}
 L = {m} \sqrt{(1 + u^0 \vec{u} \vec{v})^2-(\vec{v})^2}
 \end{equation}
Thus the canonical momenta read
 \begin{equation}
 \pi_i = \frac{\partial L}{\partial v^i} =
 \frac{{m} \left[v^i - u^i u^0 (1 + u^0 \vec{u}
 \vec{v})\right]}{\sqrt{(1 + u^0 \vec{u} \vec{v})^2-(\vec{v})^2}} =
 -m \omega_i
 \end{equation}
where we have used eq.\ \eqref{w2b}.
The Hamiltonian is
 \begin{equation}
 H = \pi_k v^k - L = \frac{{m} (1 + u^0 \vec{u}
 \vec{v})}{\sqrt{(1 + u^0 \vec{u} \vec{v})^2-(\vec{v})^2}} =
 +{m} \omega_0
 \end{equation}
Therefore the covariant four-momentum $k_\mu$ is
$ k_0 = H = {m} \omega_0, \quad
 \nad{k} = -\nad{\pi} = {m} \nad{\omega}$
i.e. $k_\mu = {m} \omega_\mu$.

Notice that
 \begin{equation}\label{k2b}
 k^2 = g^{\mu\nu}(u) k_\mu k_\nu = m^2
 \end{equation}
and the condition $H = {m} \omega_0 \ge m$
holds in each inertial frame.
The Hamilton equations for massive particle in this synchronization
have the form
 \begin{equation}\label{H1b}
 \frac{d\vec{x}}{dt} = \frac{\partial H}{\partial\nad{\pi}} = 
 - \frac{\partial k_0}{\partial\nad{k}} = \frac{\vec{k}}{k^0} =
 \vec{v}, \qquad
 \frac{d\nad{k}}{dt} = -\frac{\partial H}{\partial\vec{x}} = 0.
 \end{equation}
From the second equation it follows that $\frac{d\vec{v}}{dt} = 0$.
It is important that in this synchronization we can define
a Poincar\'e covariant Poisson structure contrary to the standard
EP synchronization case. Namely, the unique definition of
a Poisson bracket of two observables $A$ and $B$ is given by
 \begin{equation}\label{Pb2b}
 \{A, B\} = -\left({\delta^\mu}_\nu - \frac{k^\mu u_\nu}{u k}\right)
 \left(\frac{\partial A}{\partial x^\mu}
 \frac{\partial B}{\partial k_\nu}
 - \frac{\partial B}{\partial x^\mu}
 \frac{\partial A}{\partial k_\nu}\right)
 \end{equation}
with $u k = u_\mu k^\nu = u_0 k^0$; this last equality follows from
the fact that $u_k = g_{k\mu}(u) u^\mu = 0$.

It is easy to see that the Poisson bracket defined by the relation
\eqref{Pb2b} satisfies all necessary conditions:\\
--- It is linear with respect to the both factors,
      antisymmetric, satisfying the Leibniz rule and
      fulfill the Jacobi identity;\\
--- It is manifestly Poincar\'e covariant in the CT
     synchronization;\\
--- It is consistent with the Hamilton equations;\\
--- It is easy to check that the 
      dispersion relation (\ref{k2b}), $k^2 = {m}^2$,
      is consistent with this bracket
      i.e. $\{k^2, k_\nu\} = \{k^2, x^\mu\} = 0$;
      therefore we do not need to introduce a Dirac bracket.

\section{Covariant canonical formalism for the free space-like
(tachyonic) particle}

As is well known in the standard EP synchronization
in special relativity space-like trajectories are ruled out
because of causality breaking.
On the other hand, in the CT synchronization
scheme of special relativity the time component is only rescaled
by a positive factor under Lorentz group transformations.
Therefore it is possible to introduce notion of absolute causality
and consequently to overcome all difficulties of the standard
approach \cite{Rem1}.
Now, the canonical formalism for a space-like particle can be
developed in the complete analogy with the time-like case.
The corresponding action functional reads
 \begin{equation}
 S_{12}=-\kappa\int^{\lambda_2}_{\lambda_1}\sqrt{-ds^2}
 \label{S}
 \end{equation}
and under the appriopriate choice of the affine parameter
$\lambda$ we have
 \begin{equation}
 \ddot{x}^\mu=0,\qquad \omega^\mu=\dot{x}^\mu,
 \qquad \omega^2=-1
 \label{w2}
 \end{equation}
Let us focus our attention on the last constraint in the
eq.\ (\ref{w2}).
Obviously it defines an one-sheet hyperboloid;
in particular in the preferred frame
(for $u = \tilde{u} = (1, \vec{0})$)
$g_{\mu\nu}(\tilde{u}) = \eta_{\mu\nu}$, so
$\eta_{\mu\nu} w^\mu(\tilde{u}) w^\nu(\tilde{u}) = -1$,
like in the EP synchronization.
However, there is an important difference; namely under Lorentz
boosts the zeroth component $w^0(u)$ of $w^\mu$ is rescaled by a
positive factor only (see eq.\ (\ref{D:Wu})) i.e.
${w'}^0(u') = \frac{1}{W^0} w^0(u)$.
Therefore, contrary to the EP synchronization,
in this case points of the upper part of the above hyperboloid
(satisfying $w^0(u) > 0$) transforms again into
points of the upper part. This allows us to define consistently the
velocity of a tachyon:
 \begin{equation}\label{v}
 \vec{v} = \frac{d\vec{x}}{dx^0} = \frac{\vec{w}}{w^0}
 \end{equation}
because now, for each observer, the tachyon speed is finite
(i.e. $|\vec{v}| < \infty$, $w^0 > 0$). We see that the infinite
velocity is a limiting velocity, like in the non-relativistic case
(it corresponds to $w^0 = 0$ which is an invariant condition).
Notice that the constraint relation in eq.\ (\ref{w2})
implies that velocity of a tachyon moving in a direction
$\vec{n}$ is restricted by the inequality
 \begin{equation}
 |\vec{c}| = \frac{1}{1 - \vec{n} \vec{u} u^0} < |\vec{v}| < \infty.
 \end{equation}
Furthermore, the transformation law for velocities in the
CT synchronization, derived from (\ref{p1}-\ref{D:Wu}) reads
 \begin{equation}\label{v'}
 \vec{v}' = W^0 \left[\vec{v} + \vec{W} \left(\frac{(\vec{W}
 \vec{v})}{\left(1 + \sqrt{1 + (\vec{W})^2}\right)} -
 u^0 (\vec{u} \vec{v}) - 1\right)\right]
 \end{equation}
We see that the transformation law (\ref{v'}) is well defined for all
velocities (sub- and superluminal).
Recall that in the EP scheme
\emph{tachyonic velocity space does not constitute a
representation space for the Lorentz group}.
A technical point is that the space-like four-velocity
cannot be related to a three-velocity in this case by the
relation $\vec{v}_E = \frac{\vec{w}_E}{w^0_E}$, because
$w^0_E$ can take the value zero for a finite Lorentz transformation.\\
Let us identify the Lagrangian of a free tachyon related
to the action (\ref{S}); we obtain
 \begin{equation}
 L = \kappa \sqrt{(\vec{v})^2 - (1 + u^0 \vec{u} \vec{v})^2}
 \end{equation}
Thus the canonical momenta read
 \begin{equation}
 \pi_k = \frac{\partial L}{\partial v^k} =
 \frac{\kappa \left[v^k - u^k u^0 (1 + u^0 \vec{u}
 \vec{v})\right]}{\sqrt{(\vec{v})^2 - (1 + u^0 \vec{u} \vec{v})^2}} =
 -\kappa \omega_k
 \end{equation}
The Hamiltonian has the following form
 \begin{equation}
 H = \pi_k v^k - L =
 \frac{\kappa (1 + u^0 \vec{u} \vec{v})}{\sqrt{(\vec{v})^2
 - (1 + u^0 \vec{u} \vec{v})^2}} = +\kappa \omega_0
 \end{equation}
Therefore the covariant four-momentum $k_\mu$ of tachyon is
$k_\mu=\kappa\omega_\mu$

Notice that
 \begin{equation}\label{k2}
 k^2 = g^{\mu\nu}(u) k_\mu k_\nu = -\kappa^2
 \end{equation}
and the energy $H = \kappa \omega_0$ has in each inertial
frame a finite lower bound corresponding to $|\vec{v}| \to \infty$, i.e.
 \begin{equation}
 E > \frac{\kappa \sqrt{1 - (u^0)^2} \cos\phi}
 {\sqrt{1 - \left(\sqrt{1 - (u^0)^2} \cos\phi\right)^2}}
 \equiv \mathcal{E}(u^0, \phi)
 \end{equation}
where $\cos\phi = \frac{\vec{u} \vec{v}}{|\vec{u}| |\vec{v}|}$.

Therefore, contrary to the standard case,
the energy of tachyon is always restricted from below by
$\mathcal{E}(u^0, \phi) > -\infty$.
Moreover, if we calculate the contravariant four-momentum
$k^\mu = g^{\mu\nu}(u) k_\nu = \kappa \omega^\mu$ we obtain that
 \begin{equation}
 k^0 = \frac{\kappa}{\sqrt{(\vec{v})^2 - (1 + u^0 \vec{u} \vec{v})^2}} > 0
 \end{equation}
which confirm our statement that the sign of $k^0$ is
Lorentz invariant also for tachyons.
Finally, the Poisson structure can be introduced in a full
analogy with the time-like case.

\section*{Acknowledgments}
One of us (JR) is grateful to Professor H.D.~Doebner for his kind
invitaton to the XXI Internatonal Colloquium on Group Theoretical
Methods in Physics.

\end{document}